\documentclass[preprint,12pt]{elsarticle}

\usepackage{epsfig}
\usepackage{graphics}
\usepackage{graphicx}
\usepackage[centertags]{amsmath}
\usepackage{amsfonts}
\usepackage{amsthm}
\usepackage{amssymb}
\usepackage{url}
\usepackage{subfigure}
\usepackage{mhchem}
\usepackage{natbib}

\newcommand{\ra}{
\mbox{$\rightarrow$}
}

\begin{document}

\title{Computational Modalities of Belousov-Zhabotinsky Encapsulated
  Vesicles}

\author{Julian Holley, Andrew Adamatzky, Larry Bull, Ben
  {De~Lacy~Costello}, Ishrat Jahan}


\date{\today}

\begin{abstract}

\vspace{0.5cm}

\noindent

We present both simulated and partial empirical evidence for the
computational utility of many connected vesicle analogs of an
encapsulated non-linear chemical processing medium.  By connecting
small vesicles containing a solution of sub-excitable
Belousov-Zhabotinsky (BZ) reaction, sustained and propagating wave
fragments are modulated by both spatial geometry, network connectivity
and their interaction with other waves.  The processing ability is
demonstrated through the creation of simple Boolean logic gates and
then by the combination of those gates to create more complex
circuits.

\vspace{0.5cm}

\end{abstract}

\maketitle

\section{Introduction}
\label{introduction}

The last half of the twentieth century has been witness to huge leaps
in technology spanning all areas of science. One of the most
noticeable areas has been the dramatic success of the vonn
Neumann~\citep{Neumann2005} architecture electronic digital
computer. Although modern digital computers or the software has not
advanced to a point where one computer could independently create
another technologically superior computer, or where software could
compose more advanced software\footnote{A so called point of
  singularity.}, one could argue that from a purely technological
perspective such a point has already been passed. It is has now become
extremely difficult to design future computers (and develop their
software) without the aid of existing computers (and software
development tools).  In spite of such advances the current computer
architecture will always struggle with certain problems.\footnote{For
  example problems known to be NP hard, where the scale of the problem
  is known, the solution easily tested, but a solution remains
  intractable with current algorithms.}  Attempting to advance
computing beyond the current dogma, lays the field of {\it
  `Unconventional Computing'}\ \citep{UNCONVENTIONAL_2007}.  This is
an area of study that explores alternative computational
representation, substrates and strategies, the results of which not
only create novel experimental processing
devices~\citep{Adamatzky2009a}, but also contribute towards algorithms
operating on conventional serial digital computers.  One direction in
this genre is the study of reaction diffusion (RD)
computing~\citep{Adamatzky2005}.  Where the innate behaviour of a
chemical reaction and subsequent diffusion in space and time can be
used to present and manipulate information.  A suitable and convenient
chemical reaction for such processing is the Belousov-Zhabotinsky (BZ)
reaction, a type of reaction that is subject to non-equilibrium
thermodynamics creating a nonlinear chemical
oscillator~\citep{Zhabotinsky1973}.  In certain formulations the BZ
reaction can produce visible travelling waves which can be
used to represent information~\citep{Zaikin1970}.  Wave development is
effected not only by the reaction conditions, but by geometric
obstacles and collisions with other waves.  Computation {\it circuits}
analogous to electronic circuits can be created with chemical pathways
(conductors) routed through a passive substrate (insulator) with waves
representative of signals (electron flow).

In order to illustrate the possibility of computation in a BZ
substrate some of the key components that are used to create
electronic digital computers have been created, such as;
diodes~\citep{Agladze1996, Gorecka2007, Igarashia2008}, coincidence
detector~\citep{Gorecki2003} and logic gates~\citep{Toth1995,
  Steinbock1996, Motoike1999, Gorecki2009}.  These components have been combined to create more complex circuits such as
memory~\citep{Motoike2001, Gorecki2005, Gorecki2005a},
counters~\citep{Gorecki2003} and binary adders~\citep{Adamatzky2010b}.
These circuits serve to demonstrate that it is possible to create
computational devices by modelling existing digital components and
functions within the RD frame work, this approach amounts to
conventional computing on an unconventional substrate.  More
interesting are some systems that exemplify unconventional processing
and media, such as robot control~\citep{Adamatzky2004, Adamatzky2004a}
shortest path calculation~\citep{Steinbock1995, Rambidi1999,
  Rambidi2001, Agladze1997, Adamatzky2002}, image
processing~\citep{Kuhnert1989, Adamatzky2002a}, information
encoding~\citep{Bollt1997} and direction
detection~\citep{Nagahara2008}.

\begin{figure}[!ht]
\centering
\subfigure[]{\includegraphics[width=0.15\textwidth]{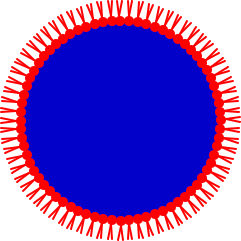}}
\hspace{10pt}
\subfigure[]{\includegraphics[width=0.28\textwidth]{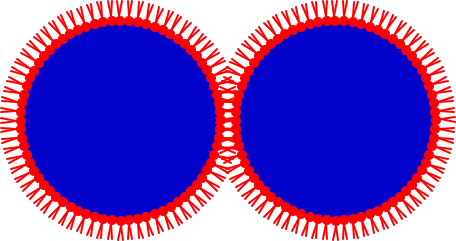}}
\hspace{10pt}
\subfigure[]{\includegraphics[width=0.3\textwidth]{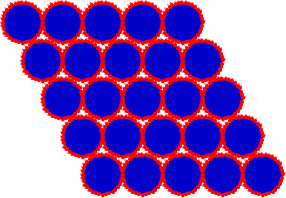}}
\caption{(a) Lipid coated mono-layer (red) vesicle enclosing BZ
  solution (blue). (b) Two vesicle lipid monolayers link to form a
  lipid bi-layer membrane connection. (c) Hexagonal array of lipid
  linked vesicles.}
\label{figure:vesicle_membrane}
\end{figure}

Contrary to these previous computation approaches in a BZ medium we
have focused on exploring the utility of connecting small spherical
processing elements containing BZ medium (vesicles) into functional
networks~\citep{neuneu2010}.  Vesicles can be created by surrounding a
solution of BZ reactant with a mono-layer of
lipids~\citep{Stanley2010}.  This cell like structure has some
interesting parallels with real neurons.  When two or more vesicles
are pressed together in solution the gap between the lipid layer forms
a chemical junction similar to a synaptic cleft.  Transmission of excitation from one vesicle to another could be possible and the
effectiveness modulated by the suspension solution.  Furthermore the
oscillatory nature of the BZ reaction can be likened to the up-state
firing (excitation) and down-state (inhibition) of neural activity.
Travelling waves can be created when the BZ solution is in a sub
excitable mode and waves can used to represent information signals,
analogous to electrical spike trains in neurons.  Connections between
vesicles could be arranged in such a way as to create functional
nuclei (Fig.~\ref{figure:vesicle_membrane}).  Reaction transmission in
mono-layer lipid coated droplets (2D vesicles) of oscillating BZ
solution has recently been reported~\citep{Szymanski2010}.

In terms of connection, self adaptation and longevity the
vesicle$\sim$neuron analogy does not hold.  Real neurons are
typified by their distributed connectedness, ability to learn and self
sustain.  Vesicles under consideration in this work are only
connected locally, also they cannot be sustained beyond exhaustion of
the reagent and at this point no adaptation mechanism has been
devised.  Nevertheless the rich phenomenological behaviour of the BZ
reaction connected in complex networks could give rise to functions
and insight to the sort of processing achieved by biological circuits.

The remainder of the paper is comprised of the following:
Section~\ref{section:simulation} details the method of BZ numerical
computer simulation and graphical presentation.
Section~\ref{section:geometry} introduces vesicle simplification,
geometry and networking.  Simulation results exploring the vesicle
geometry, connectivity and membrane function are presented in
section~\ref{section:computer_simulations}.  Elementary logic gates
are presented in section~\ref{section:logic_gates} and more complex
circuits are presented in section~\ref{section:half_adder} \&
\ref{section:memory_cell}.  The results are discussed, future
directions considered and a summary presented in the remaining
sections~\ref{section:discussion}, \ref{section:future_work} \&
\ref{section:summary}.

\section{Methods}
\label{section:methods}

\subsection{Computer simulations}
\label{section:simulation}

We have employed a two variable version of the Oregonator
model~\citep{Noyes1972} as a model of the BZ reaction~\citep{Zaikin1970,
  Zhabotinsky1973} adapted for photo-sensitive modulation of the
\cf{Ru}-catalysed reaction~\citep{Kuhnert1986}.

\begin{eqnarray}
  \frac{\partial u}{\partial t} & = & \frac{1}{\epsilon} (u - u^2 - (f
  v + \phi)\frac{u-q}{u+q})   + D_u \nabla^2 u \nonumber \\
  \frac{\partial v}{\partial t} & = & u - v \nonumber
\label{equation:oregonator}
\end{eqnarray}

Variables $u$ and $v$ are the local instantaneous dimensionless
concentrations of the bromous acid autocatalyst activator \cf{HBrO2}
and the oxidised form of the catalyst inhibitor \cf{Ru(bpy)3^3+}.
$\phi$ symbolises the rate of bromide production proportional to
applied light intensity.  Bromide \cf{Br^-} is an inhibitor of the
\cf{Ru}-catalysed reaction, therefore excitation can be modulated by
light intensity; high intensity light inhibits the reaction.  Dependant on the rate constant and reagent concentration
$\epsilon$ represents the ratio of the time scales of the two
variables $u$ and $v$.  $q$ is a scaling factor dependent on the
reaction rates alone.  The diffusion coefficients $D_u$ and $D_v$ of
$u$ and $v$ were set to unity and zero respectively. The coefficient
$D_v$ is set to zero because it is assumed that the diffusion of the catalyst is limited.

\begin{table}[tbh!]
  \small
  \centering
  \begin{tabular}{ccl}
    Parameter & Value & Description\\
    \hline\noalign{\smallskip}
    $\epsilon$ & $0.022$ & Ratio of time scale for variables $u$ and $v$ \\
    $q$ & $0.0002$ & Propagation scaling factor \\
    $f$ & $1.4$ & Stoichiometric coefficient \\
    $\phi$ & $\star$ & Excitability level (proportional to light level)\\
    $u$ & $\sim$ & Activator \cf{HBrO2}\\
    $v$ &  $\sim$ & Inhibitor \cf{Ru(bpy)3^3+}\\
    $D_u$ & $1.0$ & Activator diffusion coefficient\\ 
    $D_v$ & $0$ & Inhibitor diffusion coefficient\\ 
    $\Delta x$ & $0.25$ & Spatial step\\
    $\Delta t$ & $0.001$ & Time step\\
    \hline
  \end{tabular}
  \caption{Kinetic and numerical values used in numerical simulations
    of (Eq.~\ref{equation:oregonator}). $\star \ \phi$ Varies between
    two levels, sub-excited ($L1$) and inhibited ($L2$),
    $\phi_{L1} = 0.076$, $\phi_{L2} = 0.209$}
  \label{table:oregonator_values}
\normalsize
\end{table}

Numerical simulations were achieved by integrating the equations
using the Euler-ADI\footnote{Alternating direction implicit method.}
method~\citep{Press1992} with a time step $\delta t = 0.001$ and a
spatial step $\delta = 0.25$.  Experimental parameters are given in
Tab.~\ref{table:oregonator_values}.

Networks of discs where created by mapping 2 different $\phi$ values
(proportional to light intensity) onto a rectangle of homogeneous
simulation substrate.  To improve simulation performance the rectangle
size was automatically adapted depending on the size of the network,
but the simulation point density remained constant throughout.  The
excitation levels, $L1\ra L2$ relate to the partially active disc
interiors and non-active substrate.

Discs are always separated by a single simulation point wide boundary
layer.  Connection apertures between discs are created by
superimposing another small {\it link} disc at the point of connection
(typically a 2\ra 6 simulation point radius), simulation points have a
1:1 mapping with on screen pixels.  The reagent concentrations are
represented by a red and blue colour mapping; the activator, $u$ is
proportional to red level and inhibitor, $v$ proportional to blue.
The colour graduation is automatically calibrated to minimum and
maximum levels of concentration over the simulation matrix.  The
background illumination is mono-chromatically calibrated in the same
fashion proportional to $\phi$, white areas are inhibitory and dark
areas excited.

\begin{figure}[!ht]
\centering
\subfigure[\ra]{\includegraphics[width=0.12\textwidth]{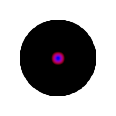}}
\subfigure[]{\includegraphics[width=0.12\textwidth]{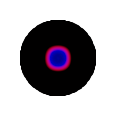}}
\subfigure[]{\includegraphics[width=0.12\textwidth]{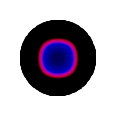}}
\subfigure[]{\includegraphics[width=0.12\textwidth]{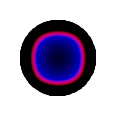}}
\subfigure[]{\includegraphics[width=0.12\textwidth]{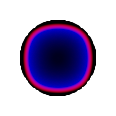}}
\subfigure[]{\includegraphics[width=0.12\textwidth]{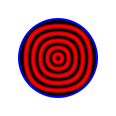}}
\caption{Example the time lapse image creation.  The image shown in
  (f) is the accumulation of successive images shown from central wave
  initiation in (a) to extinction in (e).  Time lapses periods are 50
  time steps and the refractory tail of the inhibitor ($u$) shown in
  blue is not shown in the time lapse image to improve clarity.  The
  outer blue boundary indicates an input disc when placed in a network.}
\label{figure:time_lapse_example}
\end{figure}

Wave fragment flow is represented by a series of superimposed time
lapse images (unless stated otherwise), the time lapse is 50
simulation steps. To improve clarity, only the activator ($u$) wave
front progression is recorded.  Figure~\ref{figure:time_lapse_example}
illustrates the same wave fragment in both colour map ($u$ \& $v$) and
time lapse versions ($u$).

Inputs are created by perturbing a small circular area of the
activator ($u$) set to a value of 1 with a radius of 2 simulation
points in the center of the disc.  All discs representing inputs and
outputs are highlighted with a blue and green border respectively.

\subsection{Vesicle geometry, connectivity and networking}
\label{section:geometry}

\begin{figure}[!ht]
\centering
\includegraphics[width=0.6\textwidth]{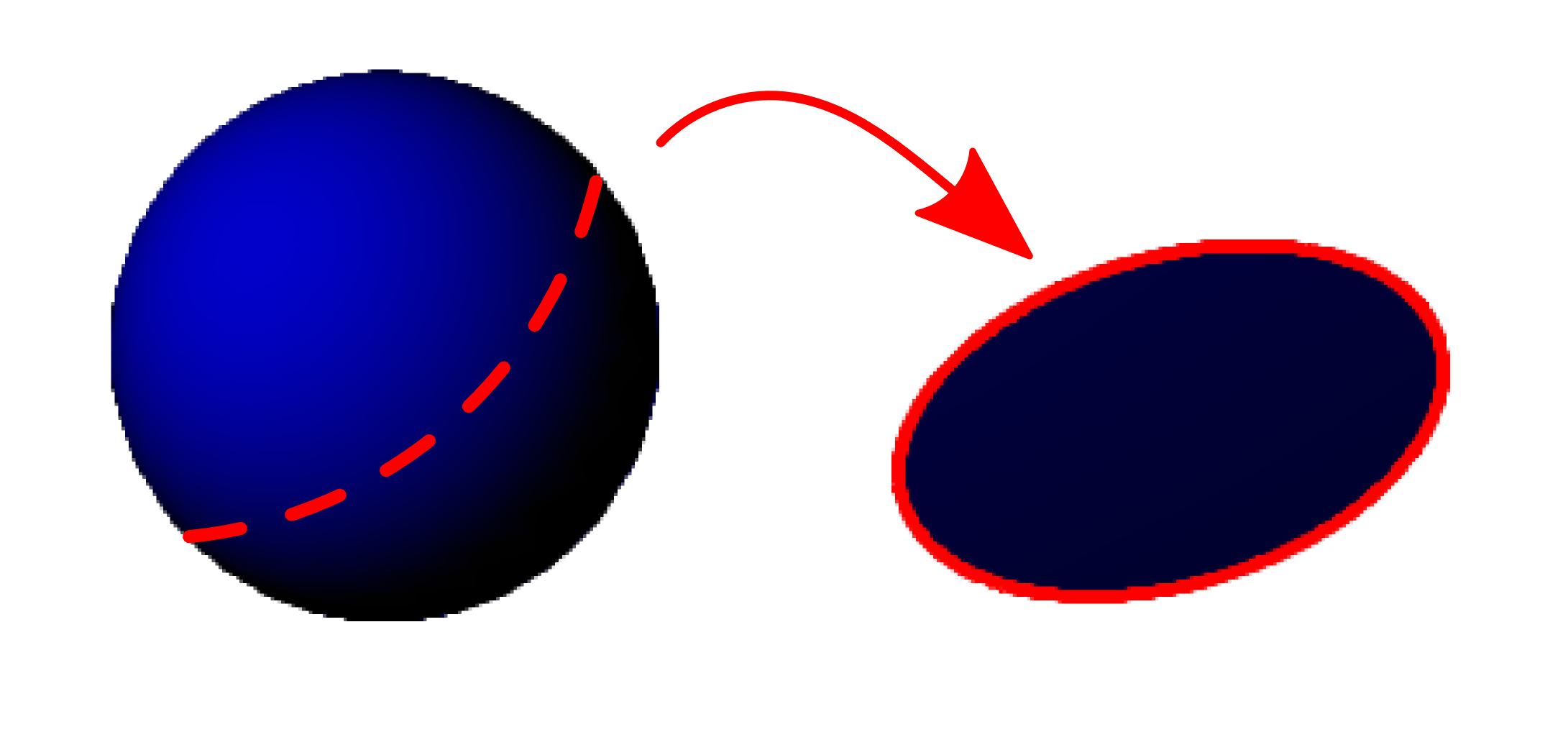}
\caption{BZ vesicle to BZ disc.  A 2D cross section of 3D vesicle of
  sub-excitable BZ medium is selected to create a {\it BZ disc}. A
  general purpose unit of connectivity and computation investigated
  during these initial explorations.}
\label{figure:vesicle_disc}
\end{figure}

The three dimensional (3D) vesicle connection opportunities and
complex internal wave reactions represent a rich computation
substrate. Such depth raises difficulties when attempting to manually
explore computation modalities.  To reduce the complexity to a level
where manual design is tractable the vesicles in this study have been
approximated into two dimensional (2D) vesicles (discs).  A disc is
created by extracting an imaginary central slice, a cross section of a
BZ vesicle (Fig.~\ref{figure:vesicle_disc}).  This reduction also
permits the opportunity of easily reproducing simulations by
projecting circuits onto a 2D photo sensitive BZ gel.  Signals are
discrete, wave fragments represent the presence or absence of a
particular signal.

\begin{figure}[!ht]
\centering
\subfigure[Relative disc sizes]{\includegraphics[width=0.3\textwidth]{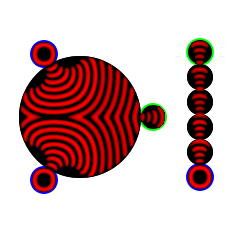}}
\subfigure[Connection angle]{\includegraphics[width=0.3\textwidth]{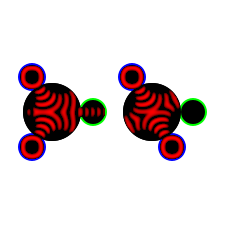}}
\subfigure[Connection efficiency]{\includegraphics[width=0.3\textwidth]{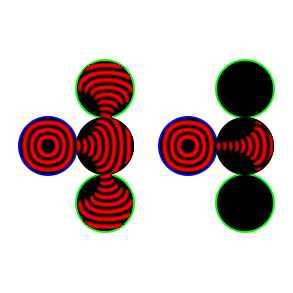}}
\caption{(a) Discs as reaction vesicles in the left network or
  communication channels in the right network. (b) The effect of
  connection angle, the two input signals combine in the left network
  to produce an output.  Adjusting the angle of the lower input in the
  right network alters the result of the collision and no output is
  produced.  (c) The effect of connection efficiency.  Large aperture
  (6 points) connection in the left network results in a broad
  spreading beam. Conversely a smaller (4 point) aperture connection
  in the right network creates a narrow beam wave.  (In all images,
  inputs occur in discs circled in blue, outputs circled in green.)}
\label{figure:signal_modulation}
\end{figure}

In a previous study we have shown that logic circuits can be created
with uniform discs arranged in hexagonal
networks~\citep{Adamatzky2010b}, hexagonal packing being the most
efficient method of sphere (disc) packing.  Further opportunities to
modulate wave fragment behaviour are presented when disc size,
connection angle and connection efficacy are combined in
non-homogeneous networks.  Disc size can be adjusted to permit or
restrict internal wave interactions, producing either larger reaction
vessel discs or smaller communications
discs~(Fig.~\ref{figure:signal_modulation}a).  Connection angle
between discs can be used to direct wave
collisions~(Fig.~\ref{figure:signal_modulation}b) and connection
efficiency can effect the wave
focus~(Fig.~\ref{figure:signal_modulation}c).

\subsection{Computer simulation experiments}
\label{section:computer_simulations}

\begin{figure}[!ht]
\centering
\includegraphics[width=0.8\textwidth]{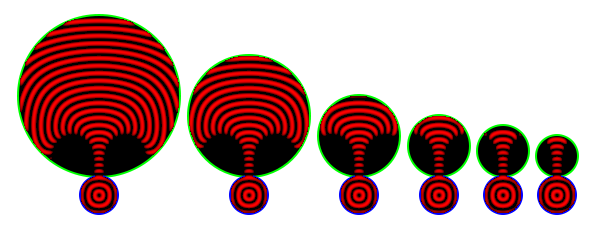}
\caption{Wave fragment propagating from an fixed lower
  input disc of size 18 units to variable upper larger discs of radius
  80 (left), 60, 40, 30, 25 \& 20 (right) units.  Source disc has a
  radius of 18 points and a connection aperture of 4 units.}
\label{figure:disc_size_demo}
\end{figure}

Increasing the relative disk size can be used not only to allow space
for wave fragment collisions~(Fig.~\ref{figure:signal_modulation}a)
but other effects are also apparent.
Figure~\ref{figure:disc_size_demo} illustrates the front development
of the same wave through progressively smaller terminating discs.  In
the larger discs the wave fragment has more space in which to develop
and spreads out to the majority of the disc perimeter, conversely
in the smaller discs the wave fragment doesn't have time to develop and
terminates almost directly opposite the entry aperture.

Wave fragments cannot survive when the fragment mass drops below some
critical level~\citep{Kusumi1997} and this is evident when comparing
progressively smaller aperture sizes with fixed size discs.  In our
system and with a disc radius of 28 units\footnote{Simulation grid
  points.} the critical level surrounds an aperture gap of 4 units.
Below that fragments do propagate through the aperture but
quickly die.  The narrow beam produced as a result of a 4 unit
aperture (type J1) presents an opportunity to deflect the wave to
alternate exits and perform ballistic style
computation~\citep{Fredkin2002}.  We have found that using a narrow
beam aperture in orthogonal networks where wave fragments do not
normally propagate into connected perpendicular discs particularly
useful in creating simple logic gates
(Sect.~\ref{section:logic_gates}).  Furthermore, diode junctions can
be created when networks of narrow (type J1) and broadband (type J2)
are combined~(Sect.\ref{section:half_adder}).  Although more
functionality could be achieved with more subtle aperture
adjustments~\citep{Adamatzky2010c} further explorations in this work rely on
combining just the two types, narrow (J1) and broadband (J2)
(Fig.~\ref{figure:signal_modulation}c).

\begin{figure}[!ht]
\centering
\subfigure{\includegraphics[width=0.6\textwidth]{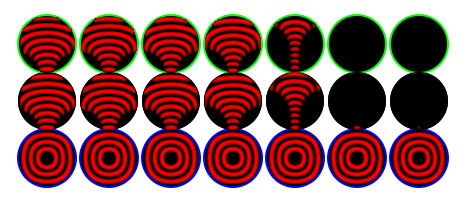}}
\caption{Comparison of aperture sizes.  All disc sizes remain fixed at
  28 units whilst the aperture is reduced from 8 (leftmost) to 2 units
  (rightmost).}
\label{figure:aperature_array}
\end{figure}

\section{Results}
\label{section:results}

\subsection{Elementary logic gates}
\label{section:logic_gates}

\begin{figure}[!ht]
\centering
\subfigure[0\ra 1]{\includegraphics[width=0.3\textwidth]{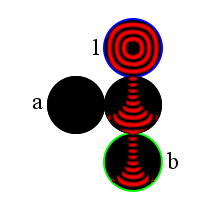}}
\subfigure[1\ra 0]{\includegraphics[width=0.3\textwidth]{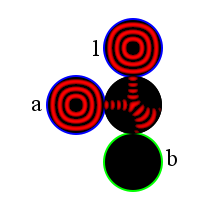}}
\caption{Inverter gate ($a = \overline{b}$) where the input ($a$) is
  center left (blue ring), bottom disc (green ring) is the output
  ($b$) and a supply, or source logical `1' top most disk (blue ring).
  (a) $a = 0$ The gate initiates with the source pulse in the top
  disc.  In this case, no signal is present at the input disc and the
  source pulse travels to the output disc (bottom) resulting in a
  logical 1 output (1\ra 0). (b) $a = 1$ Again the source pulse
  travels from top to bottom, but in this case a collision with a
  signal present on the input disc produces a logical 0 output. (0\ra
  1).}
\label{figure:inverter_gate}
\end{figure}

Electronic logical gates form the building blocks of more complex
digital circuitry forming the foundations of complex high level
components such as microprocessors.  Although we do not envisage
creating traditional vonn Neumann architecture microprocessors in BZ
vesicles, the ability to create simple logic gates with BZ vesicles
demonstrates that (like electronics) the medium and architecture is
{\it capable} of such processing.  Logic gates and composite circuits
of logic gates have been created several times before using the BZ
substrate, for example~\citep{Toth1995, Steinbock1996,
  Motoike1999, Gorecki2009}. Here we illustrate a selection of key
gates can be created using nothing other than interconnected BZ discs.
Figure~\ref{figure:inverter_gate} illustrates the operation of the
most elementary of gates the inverter (NOT gate).  The circuit
operation starts with the simultaneous application of the circuit
input (left most disc) in conjunction with the source (permanent
logical `1') input (top most disc).  The circuit operation terminates
by observing the output disc (lower most disc) at a time when either
result state would be present.  If the progression of a wave fragment
through a disc is considered as 1 step then the output disc will hold
a valid result after 2 steps from the application of the source input.
Incorporating a parallel un-modulated source signal that travels from
output to input could also be used to indicate the point at which the
output discs holds a valid output.  In this case this would simply
consist of 3 serial discs.

\begin{figure}[!ht]
\centering
\subfigure[01\ra 0]{\includegraphics[width=0.3\textwidth]{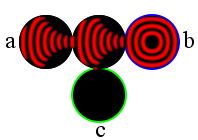}}
\subfigure[10\ra 0]{\includegraphics[width=0.3\textwidth]{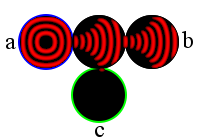}}
\subfigure[11\ra 1]{\includegraphics[width=0.3\textwidth]{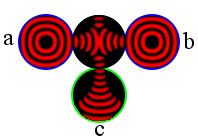}}
\caption{Two input AND gate ($c = a\bullet b$) where inputs $a,b$ are
  top left and right discs (blue rings) and output $c$ is the bottom
  central disc (green ring). (a) $(a,b)(0,1)$ A wave from input $b$
  propagates uninterrupted and terminates in the opposing input disc
  $a$. (b) $(a,b)(1,0)$ Likewise, a wave from input $b$ propagates
  uninterrupted and terminates in the input disc $b$. (c) $(a,b)(1,1)$
  Waves from both input discs $a$ and $b$ collide in the central disc
  and eject two perpendicular waves, one of which propagates into the
  output disc ($c$).}
\label{figure:and_gate}
\end{figure}

\begin{figure}[!ht]
\centering
\subfigure[00 \ra 1]{\includegraphics[width=0.36\textwidth]{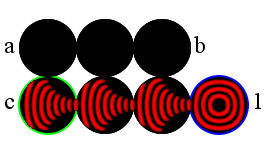}}
\subfigure[01 \ra 1]{\includegraphics[width=0.36\textwidth]{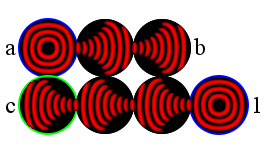}}
\subfigure[10 \ra 1]{\includegraphics[width=0.36\textwidth]{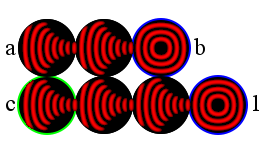}}
\subfigure[11 \ra 0]{\includegraphics[width=0.36\textwidth]{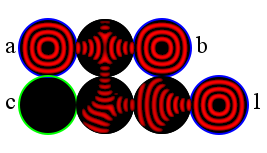}}
\caption{Two input NAND gate ($c = \overline{a\bullet b}$) where
  inputs $a,b$ are top left and right discs (blue rings) and output
  $c$ is the bottom left disc (green ring), source input is located on
  the bottom right (blue ring).  Operation is identical to the AND
  gate (Fig.~\ref{figure:and_gate}) but with an inverter
  (Fig.~\ref{figure:inverter_gate}) integrated along the bottom disc
  row. (a), (b) \& (c) The source input provides a logical `1' output
  for all input combinations other than $(a,b)(1,1)$. (d) $(a,b)(1,1)$
  Output from the AND gate portion of the gate collides with the
  source input creating a logical `0' output. }
\label{figure:nand_gate}
\end{figure}

The operation of an AND gate and the inversion, the NAND gate are shown
in Fig.~\ref{figure:and_gate} \& Fig.~\ref{figure:nand_gate}.  The
result of a wave collision in the NOT gate was exploited to deflect
and extinguish the source wave into the disc edge, whereas in the AND
gate the collision between the two inputs results in 2 perpendicular
fragments, one of which develops in the output cell to produce the
result.

A NAND gate can created by combining the NOT gate and the AND
gate~Fig.\ref{figure:nand_gate}.  NAND gates are known as {\it
  universal} gates since all other gates can be created from
arrangements of NAND gates alone.\footnote{NOR gates are also
  universal gates.}  The NOT gate (Fig.~\ref{figure:inverter_gate}) is
integrated below the AND gate in the lower row
(Fig.~\ref{figure:and_gate}) where the activity of a horizontal source
signal inverts the vertical output.

\begin{figure}[!ht]
\centering
\subfigure[00 \ra 0]{\includegraphics[width=0.3\textwidth]{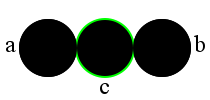}}
\subfigure[01 \ra 1]{\includegraphics[width=0.3\textwidth]{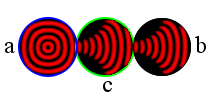}}
\subfigure[10 \ra 1]{\includegraphics[width=0.3\textwidth]{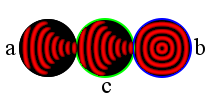}}
\subfigure[11 \ra 1]{\includegraphics[width=0.3\textwidth]{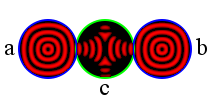}}
\caption{Two input OR gate ($c = a+b$) where inputs $a,b$ are left and
  right discs (blue rings) and the central disc is the output $c$
  (green ring).}
\label{figure:or_gate}
\end{figure}

The OR gate is used to detect the presence of one or more signals.  A
logical `1' on any input results in an output
(Fig.~\ref{figure:or_gate}).  Common amongst all these gates, the
output value of a logical `1' or `0' as indicated by the presence or
absence of wave is only valid at a specific point in the development and
in these instances, approximated to be proportional to time.  For
example the OR gate output is sampled after a wave fragment has
travelled by one disc unit ($t_d = 1$).  Therefore the annihilation of
the ($a,b$)($1,1$) case and the continuation of the waves into
opposing input cells for cases ($a,b$)($0,1$) and ($a,b$)($1,0$) does
not effect the outcome.

\begin{figure}[!ht]
\centering
\subfigure[00 \ra 1]{\includegraphics[width=0.3\textwidth]{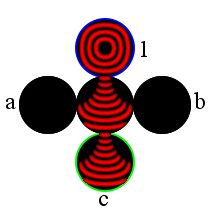}}
\subfigure[01 \ra 0]{\includegraphics[width=0.3\textwidth]{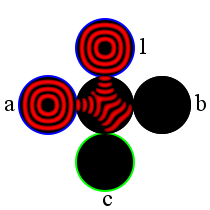}}
\subfigure[10 \ra 0]{\includegraphics[width=0.3\textwidth]{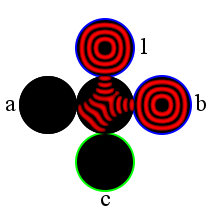}}
\subfigure[11 \ra 1]{\includegraphics[width=0.3\textwidth]{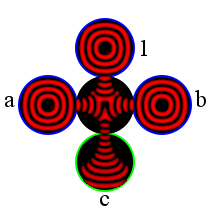}}
\caption{Two input NXOR gate ($c = \overline{a\oplus b}$) where inputs
  are middle row left and right (blue rings), the output disc is
  center bottom (green ring) and source input center top (blue ring).
  The OR structure (Fig~\ref{figure:or_gate}) is repeated in the
  central row, the output of which deflects the source input from the
  center top disc, the result is an NXOR gate.  The output of the NXOR
  can then be inverted to create a XOR gate
  (Fig.~\ref{figure:xor_gate}).}
\label{figure:nxor_gate}
\end{figure}

\begin{figure}[!ht]
\centering
\subfigure[00\ra 0]{\includegraphics[width=0.36\textwidth]{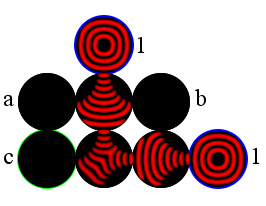}}
\subfigure[01\ra 1]{\includegraphics[width=0.36\textwidth]{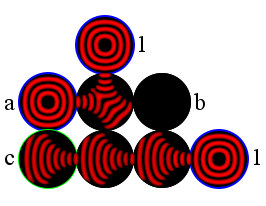}}
\subfigure[10\ra 1]{\includegraphics[width=0.36\textwidth]{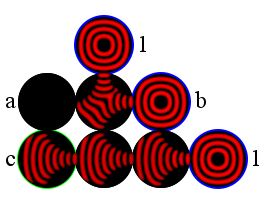}}
\subfigure[11\ra 0]{\includegraphics[width=0.36\textwidth]{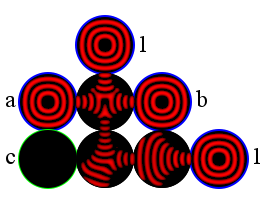}}
\caption{Two input XOR gate ($c = a\oplus b$) where inputs are middle
  row left and right (blue rings), the output disc is bottom left
  (green ring) and there are two source inputs, center top and bottom
  right (blue rings).  The gate is an extension of the inverting the
  NXOR gate (Fig.~\ref{figure:nxor_gate}).}
\label{figure:xor_gate}
\end{figure}

The XOR gate is used to signal a difference between signals, producing
an output when inputs alternate regardless of the composition of the
difference.  Figure~\ref{figure:xor_gate} illustrates the BZ disc
implementation along with the inversion NXOR in
Fig.~\ref{figure:nxor_gate}.

\subsection{Half adder}
\label{section:half_adder}

\begin{figure}[!ht]
\centering
\subfigure[One bit half adder]{\includegraphics[width=0.3\textwidth]{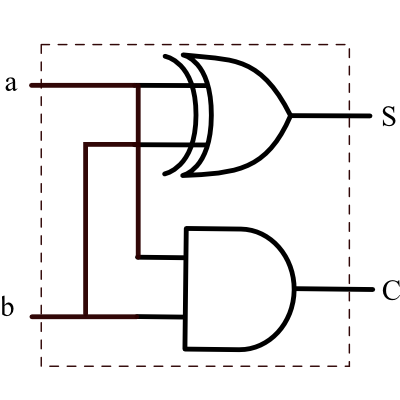}}
\hspace{10pt}
\subfigure[One bit full adder]{\includegraphics[width=0.5\textwidth]{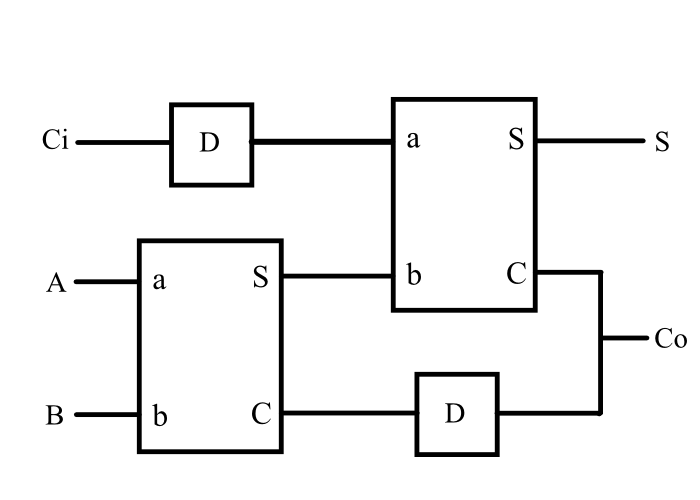}}
\caption{(a) One bit half adder.  The circuit comprises of two
  outputs, the sum ($S$) derived from the XOR gate ($S = a\oplus b$)
  and carry ($C$) derived from the AND gate ($C = a\bullet b$).  (b)
  One bit full adder.  Two half adders can be cascaded together to
  create a full 1 bit adder ($S = C_i\oplus(a\oplus b)$, $C_o =
  a\bullet b + (a\oplus b)$).  In turn full 1 bit adders can be
  cascaded to create an $N$ bit adder. The $D$ blocks represent signal
  delays required in order to synchronise signal pulses from different
  sources.}
\label{figure:ha_circuit}
\end{figure}

The half adder is a sub-system used in binary addition circuits.  The
half adder adds two binary digits and when connected with another half
adder creates a full 1 bit adder.  One bit adders can then in turn be
connected together to make $n^{th}$ bit adders
(Fig.~\ref{figure:ha_circuit}).  A half adder can be constructed from
a combination of two logic gates the XOR and AND gate.  There are two
inputs ($a$ \& $b$) and two outputs ($S$ \& $C$), the binary sum ($S$)
of $a$ \& $b$ is achieved by the XOR gate ($S = a \oplus b$) and
inability of the configuration (overflow) to present the $1 + 1$ input
is achieved with a carry ($C$) output, ($C = a\bullet b$).

\begin{figure}[!ht]
\centering
\subfigure[$ab(0,0)$\ra $S = 0,\ C = 0$]{\includegraphics[width=0.4\textwidth]{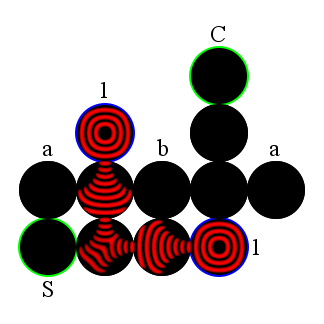}}
\subfigure[$ab(1,0)$\ra $S = 1,\ C = 0$]{\includegraphics[width=0.4\textwidth]{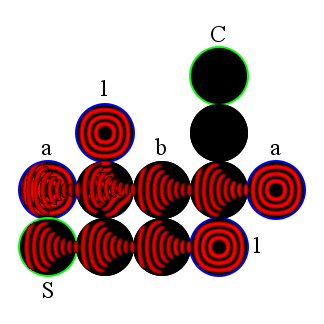}}
\subfigure[$ab(0,1)$\ra $S = 1,\ C = 0$]{\includegraphics[width=0.4\textwidth]{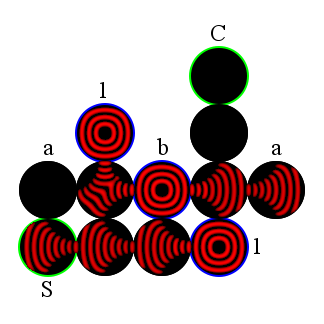}}
\subfigure[$ab(1,1)$\ra $S = 0,\ C = 1$]{\includegraphics[width=0.4\textwidth]{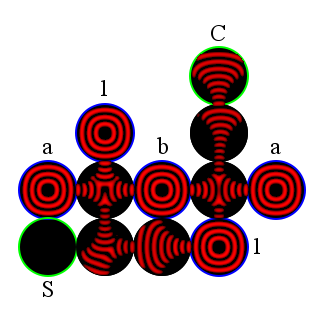}}
\caption{Half adder circuit ($S = a\oplus b,\ C = a\bullet b$) where
  inputs are located along the central row, $a$ far left, $b$ central
  and then $a$ repeated far right (blue rings), two source inputs
  located top left and bottom right (blue ring).  The output cell is
  bottom left (green ring). The circuit is constructed by combining an
  XOR gate (Fig.~\ref{figure:xor_gate}) (bottom left) and the AND gate
  (Fig.~\ref{figure:and_gate}) (top right).  The issue of the signal
  passing problem is obviated by replicating one of the inputs `$a$' (see {\it signal passing problem}
  below).}
\label{figure:ha_gate}
\end{figure}

\begin{figure}[!ht]
\centering
\subfigure[]{\includegraphics[width=0.32\textwidth]{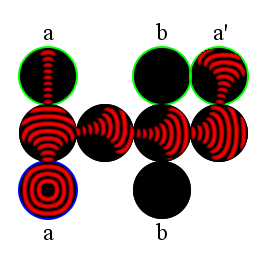}}
\subfigure[]{\includegraphics[width=0.32\textwidth]{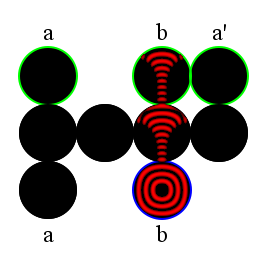}}
\subfigure[]{\includegraphics[width=0.32\textwidth]{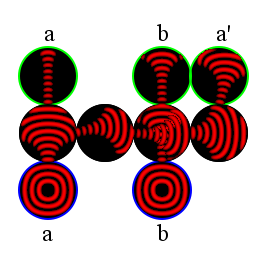}}
\caption{Signal passing problem resolved by sharing the same space (a
  {\it cross road}) but at different times.  Signals travel from
  bottom to top. (a) Shows the independent path of signal $a$.  The
  signal is split by using a broad band aperture at the junction
  between the $1^{st}$ and $2^{nd}$ disc to create $a'$. (b) Shows the
  independent path of signal $b$.  (c) Shows $a'$ crossing $b$.}
\label{figure:signal_passing}
\end{figure}

\begin{figure}[!ht]
\centering
\subfigure[]{\includegraphics[width=0.26\textwidth]{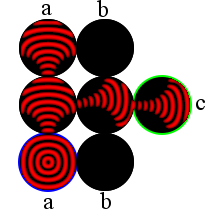}}
\subfigure[]{\includegraphics[width=0.26\textwidth]{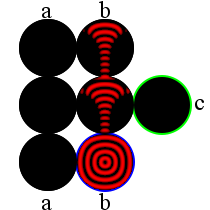}}
\subfigure[]{\includegraphics[width=0.26\textwidth]{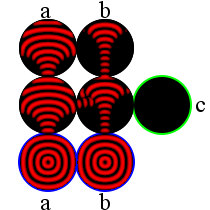}}
\caption{Signal passing gate.  When the time difference between two
  signals trying to cross is too small the refractory tail of one
  signal will cause extinction in the other.  This feature can be
  exploited to create another logic gate.  The output in this gate
  produces $c = a\bullet\overline{b}$. (a) Shows the independent path of
  signal $a$.  Signal $a$ is split by using a broad band aperture at
  the junction between the $1^{st}$ and $2^{nd}$ disc to create
  $a'$. (b) Shows the independent path of signal $b$.  (c) The
  refractory tail of $b$ extinguishes signal $a'$.}
\label{figure:signal_passing_gate}
\end{figure}

A 1 bit half adder created from BZ discs can also be constructed from
connecting a BZ disc AND gate and XOR gate
(Sect. \ref{section:logic_gates}).  Figure~\ref{figure:ha_gate} shows
the BZ disc conjunction for the half adder circuit.  The input $a$
needs to be repeated on the other side of input $b$ in order for this
circuit to work.  This is necessary in order to overcome the {\it
  signal passing problem}, a universal problem for systems where
signals propagate along specific planular channels.  There are two
ways to overcome this problem, either add identity to the signals in
such a way that signals can share the medium or share the medium at
different times.  How two or more waves could be identified and share
the same space in this BZ system remains unclear because of the diffusive
nature of the reaction.  However sharing a channel medium in
time\footnote{In communications systems this is known as Time Division
  Multiplexing (TDM).} is possible if the time difference between
signals is large enough to prevent the refractory tail from one
extinguishing the other.  Figure~\ref{figure:signal_passing}
illustrates one such temporal separation strategy, where signal $a$
passes over signal $b$ but becomes shifted in time.  The circuit
operates with two types of apertures, one that creates a narrow beam
(type J1) wave and another that creates a broad beam (type J2) wave.
Signals $a$ \& $b$ travel from bottom to top, with $a$ on the left and
$b$ on the right.  The signal $a$ is split at the junction to the
first disc and a fragment $a'$ travels horizontally towards
$b$. Meanwhile $b$ is already traversing the first disc and has
progressed into the final disc before $a'$ crosses the $b$ path
allowing $a'$ to cross $b$.  A time shift $t_d$ now exists between
$a,b$ and $a',b$ so any further processing between $a'$ and $b$ must
therefore delay $b$ by $t_d$.  This strategy relies on allowing
sufficient time for the refractory tail of signal $b$ to have a
negligible effect on $a$.  If the signals are not sufficiently
separated then $b$ will extinguish $a'$ which can in another context
be used as another logical construction
(Fig.~\ref{figure:signal_passing_gate}).

Venturing into 3 dimensions (3D) resolves the signal passing problem
all together, allowing signals to be routed vertically.  At this stage
only 2 dimensional (2D) structures of discs have been explored, but
these are approximations of our target computation node, a 3D BZ
vesicle.  In this current 2D perspective, overcoming the signal
passing problem via interconnecting linking layers above and below
planular 2D functions seems the next logical step analogous to a
methodology used in 2 layer and multilayer electronic circuit boards.

\begin{figure}[!ht]
\centering
\subfigure[$ab(1,0)$\ra $S = 1,\ C = 0$]{\includegraphics[width=0.34\textwidth]{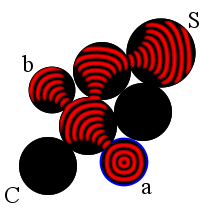}}
\subfigure[$ab(0,1)$\ra $S = 1,\ C = 0$]{\includegraphics[width=0.34\textwidth]{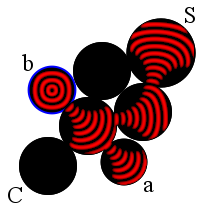}}
\subfigure[$ab(1,1)$\ra $S = 0,\ C = 1$]{\includegraphics[width=0.34\textwidth]{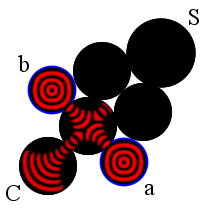}}
\caption{Composite half adder circuit ($S = a\oplus b,\ C = a\bullet
  b$) where inputs and outputs are all connected to a central reactor
  disc which can achieve both the AND and XOR function.  The two {\it
    half} outputs from the XOR operation are recombined with an OR
  operation with addition discs in the top right.  The circuit uses
  all 3 methods of modulation, connection angle, disc size and
  aperture efficacy (Sect.~\ref{section:geometry}).}
\label{figure:ha_gate2}
\end{figure}

Another specific solution for the half adder circuit which
removes the need to repeat one of the inputs is possible if all the
signal modulation techniques are exploited; disc connection geometry,
disc size and aperture efficacy (Sect.~\ref{section:geometry}).
Figure~\ref{figure:ha_gate2} demonstrates a half adder design where
most of the processing occurs in one central reactor disc.  The
central disc achieves the AND function (Fig.~\ref{figure:ha_gate2}c)
and the XOR function (Fig.~\ref{figure:ha_gate2}b \& c).  Considering
the central disc principally in terms of an AND gate; then the XOR
function can be considered as being derived from the AND gate response
to input sets $(a,b)(0,1)$ and $(a,b)(1,0)$.  The outputs of which are
curved around into an OR gate in the $S$ output disc creating the
XOR function.

\clearpage
\subsection{Memory cells}
\label{section:memory_cell}

Memory is an essential facet of both adaptive behaviour in Nature and
in synthetic computation.  It permits animals and machines to build an
internal state independent from the current external world state.  In
this section we present a simple 1 bit volatile read write memory cell
constructed entirely with BZ discs.  The cell design is independent
but similar to previous designs~\citep{Motoike1999, Motoike2001} in so
much that the existence or absence of a rotating wave represents the
setting or resetting of 1 bit of information.

\begin{figure}[!ht]
\centering
\subfigure[]{\includegraphics[width=0.3\textwidth]{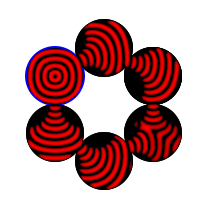}}
\subfigure[]{\includegraphics[width=0.3\textwidth]{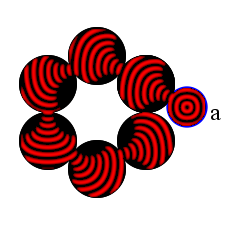}}
\subfigure[]{\includegraphics[width=0.3\textwidth]{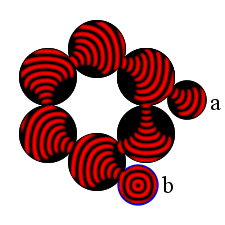}}
\caption{Memory cell development. (a) Hexagonal connected loop of
  discs, perturbing the medium in any of the cells always leads to
  complete wave extinction as counter-clockwise and clockwise wave
  fronts meet at some point during the path. (b) Insertion of a angled
  {\it diode} junction inserts a uni-direction (counter-clockwise)
  wave that rotates indefinitely (in simulation).  (b) Addition of
  another opposing diode junction provides for the insertion of
  uni-directional (clockwise) wave.  Insertion of a wave from either
  input disc (a|b) can be cancelled by inserting another asynchronous
  uni-directional wave from the other opposing disc.  Inserting more
  than one wave is not sustainable and always results in a reduction
  to one wave front.}
\label{figure:memory_cell}
\end{figure}

When two BZ waves progress in opposite directions around an enclosed
channel, loop or ring of connected discs, then at some point the two
opposing wave fronts will meet and are always mutually annihilated
(Fig.~\ref{figure:memory_cell}a).  Nevertheless, if a unidirectional
wave can be inserted into the loop then that wave front will rotate
around the loop indefinitely\footnote{For as long as the chemical
  reagents can sustain the reaction.}
(Fig.~\ref{figure:memory_cell}b).  Furthermore the rotating wave can
be terminated by the injection of another asynchronous wave rotating
in the opposite direction (Fig.~\ref{figure:memory_cell}.c).  Opposing
inputs into a loop are analogous to a memory {\it set} or {\it reset}.
Reading the state of the cell without changing the state can be
achieved by connecting another output node where a stream of pulses
can be directed to modulate other circuits~\citep{Gorecki2003a}.

\begin{figure}[!ht]
\centering
\subfigure[Pass $a$\ra $b$]{\includegraphics[width=0.26\textwidth]{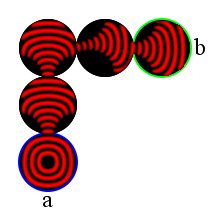}}
\subfigure[Block $b$\ra $a$]{\includegraphics[width=0.26\textwidth]{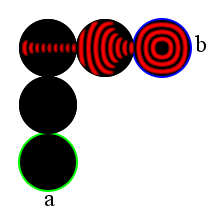}}
\caption{BZ disc diode. (a) Signal propagates from bottom to top
  ($a\ra b$).  A broad band (type J2) aperture at the $2^{nd}$ (right
  angle) junction connection permits the signal to expand horizontally
  towards $b$. (b) Signal propagates from right to left ($b\ra a$).
  Conversely the narrow band junction at the $2{nd}$ (right angle)
  junction prohibits propagation towards $a$.}
\label{figure:right_angle_diode}
\end{figure}

The loop and a unidirectional gate (diode) are the two key
constructions of this type of memory cell.  Unidirectional gates in BZ
media have previously been created by exploiting asymmetric geometries
or chemistry on either side of a barrier~\citep{Agladze1996}.  An
alternative design is possible however using discs connected with
different apertures.  Figure~\ref{figure:right_angle_diode}
illustrates a diode constructed from a right angle junction connected
by a broad band (type J2) aperture to a vertical column and by a
narrow beam (type J1) aperture to a horizontal row.  Signal flow is
only possible from bottom to top ($a\ra b$) because of the asymmetric
apertures in the right angle connecting the disc.  The operation relies on
the relationship between the wave expansion and the angle of
the connection.  Fine control of the wave beam would in theory allow
other angles of connectivity~\citep{Adamatzky2010c} and other
functions.  In practice fine control of wave diffusion is however
difficult to achieve and hence we have restricted our choice between
just two types.

\begin{figure}[!ht]
\centering
\subfigure[]{\includegraphics[width=0.4\textwidth]{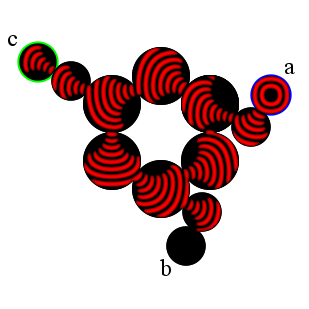}}
\subfigure[]{\includegraphics[width=0.4\textwidth]{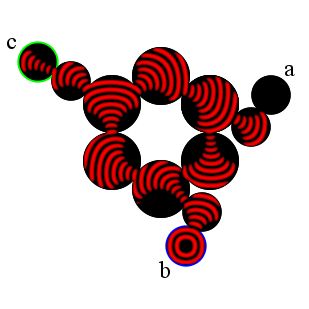}}
\subfigure[]{\includegraphics[width=0.4\textwidth]{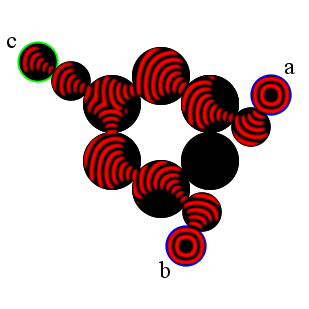}}
\caption{Memory cell with additional diodes on the cell inputs.  Two
  additional angled diode junctions are added to each of the input
  discs ($a$ \& $b$).  This prevents a reverse wave flow back down
  either of the inputs.  An example output disc is also connected
  (top left).  (a) Wave insertion at (top right) $a$ input node
  results in a persistent counter-clockwise wave.  Reverse wave flow
  down the opposing (bottom) input is blocked by an angled diode
  junction.  (b) Wave insertion at (bottom) $b$ input node results in
  a persistent clockwise wave.  Reverse wave flow down the opposing
  (top right) $a$ input is likewise, blocked by an angled diode
  junction. (c) Simultaneous $a$ \& $b$ inputs produce one output
  pulse ($c$) and annihilate wave rotation.}
\label{figure:memory_cell2}
\end{figure}

As the rotating wave progresses around the loop in the memory cell
illustrated in Fig.~\ref{figure:memory_cell}, the opposing input cell
also inadvertently becomes an output cell.  This may be undesirable in
some designs but can be easily resolved by adding another pair of
diode junctions to the circuit. Figure~\ref{figure:memory_cell2} shows
such a design, where opposing inputs are not affected by the opposing
input.

\clearpage
\section{Discussion}
\label{section:discussion}

Our research is an exploratory component within a collaborative
project that aims to create a lipid encapsulated BZ vesicle and
organise those vesicles into a functional network.  The lipid membrane
and the non-linear oscillatory nature of the BZ medium encodes some of
the features apparent in biological information processing.  Whilst
inter-neuron communication is electrical, modulation of that activity
is chemical and in the case of individual neurons modulation dominates
at the synaptic junction.  The membrane between two vesicles can be
considered a simple analog of the synaptic junction, a small contact
area that can modulate signals in between vesicles.  Similarly the
electric upstate firing and downstate quiescence of neural signalling
is an analog of chemical excitation and refraction.  Connecting
vesicles together presents the fascinating possibility of creating a
chemical processing device similar in principle to biological
systems.~\citep{neuneu2010}

Another analog between Natural processing and vesicles is the
relationship to artificial Life.  The {\it cell} is the building block
of all known life on Earth.  Mechanistic explanation for the genesis
of Life and the cell remain elusive, but a key aspect of cell
morphology is the concept of a cell wall and the ability to separate
one state (the outside) from another (the inside), the possibility for
an increase in entropy~\citep{Harold2001}. One theory (amongst many)
is a role for lipids in the spontaneous formation of simple cells and
hence the development of a separate entity different from the
surrounding environment.  Whether spontaneous lipid cell formation
played a role in early Life remains to be seen, but the principle of
an enclosing membrane to a cell like unit appears essential in order
separate environment from agent.  Enclosing a nonlinear chemical
oscillator such as the BZ reaction into a cell (vesicle) leads to
another type of phenomena.  If the reaction (upstate) in a vesicle can
in some degree migrate across the membrane then the reaction in one
vesicle could influence the reaction in another, and in turn be
subject of influence.  The nature of the excited and refractory
temporal dynamics of the reaction can lead to interesting emergent
ensemble behaviour~\citep{Vanag2001}.  Computer simulations of such
behaviour of similar simple processing units, known as {\it `Cellular
  Automata'} (CA) has been extensively studied~\citep{Wolfram2002} and
can lead to interesting Life like behaviour~\citep{Berlekamp1982}.

The exploration in the computation modalities of BZ encapsulated
vesicles is promising then on (at least) two levels.  The macro scale
of organised behaviour (classification of this work) and the small
scale emergent oscillatory behaviour analogous to cellular automata.
The parallel between the notion of a conscious single thread behaviour
and the unconscious parallel emergent behaviour.  In this study we
have shown that most of the computation accomplishments of previous
geometrically constrained BZ processing at the macro scale can be
achieved with BZ discs alone. In extrapolating discs into spherical
vesicles more interesting behaviour and processing is likely to be
possible\footnote{The innate resolution of the signal passing problem
  (Sect.\ref{section:half_adder}) for example.} albeit at the cost of
the simplicity and clarity of design.

\section{Future work}
\label{section:future_work}

Experiments are currently in progress to replicate these simulation
results in real chemistry (an example of the AND gate is shown in the
appendix (Sect.~\ref{section:appendix})).  Our goal is to explore
computational modalities of interconnected discs and vesicles.  In
doing so we hope that such work will both inspire novel chemical
computing and introduce new strategies for use in existing systems and
other mediums (including silicon) or conventional computers.
Implementing devices that are known to be an essential to perform
conventional computation is useful for demonstrative purposes.
Expunging the computing capabilities of geometrically modulated
reaction diffusion computers. Nevertheless the innate massively
parallel and deep temporal-spatial nature of such a substrate is a
good candidate to explore the kind of computation problems for which
vonn Neumann architecture machines perform so poorly.  Previous
studies have shown that RD systems are capable of performing
computation in the conventional paradigm and this study has shown that
distributing encapsulated RD units in cell like units is equally
competent.  On that basis other perhaps unknown structures and
strategies could be developed beyond our current understanding.  To
achieve these aims and in part, not to be biased by known solutions
and tradition design methodology our next step is to apply an
evolutionary strategy to evolve functional networks of discs.  We
intend to focus on solving computational tasks for which the solutions
are currently protracted in conventional systems.

\section{Summary}
\label{section:summary}

Creating components, gates and circuits commonly used in the design of
discrete conventional computers within geometrically constrained
constructions containing sub-excitable BZ media has been extensively
studied both in simulation and real chemistry.  This work has shown
that some of the previous circuits, logic gates, composite logic gate
circuits (the half adder) and memory can be reproduced using networks
of interconnected discs alone.  Wave modulation through discs can be
manipulated by changing the network interconnections, relative disc
sizes and aperture efficiency.  All the designs presented rely on a
uniform excitability level of the reaction.  This is an important
consideration since discs or vesicles whose function relies on
non-uniform excitation levels may eventually fail as reagents equalise
across connecting junctions, additionally photo modulation of
excitability is possible with discs in 2 dimensions, but otherwise
impossible with vesicles in 3 dimensional structures. Elementary and
universal logic gates and composite circuits have also been shown
possible with a uniform disc size and aperture gap connected in a
simple orthogonal network structure, whereas other circuits such
as the diode and memory cell rely on a combination of different
geometric connectivity, disc size and aperture efficacy.

\clearpage
\section{Appendix}
\label{section:appendix}

\begin{figure}[!ht]
\centering
\subfigure[]{\includegraphics[width=0.4\textwidth]{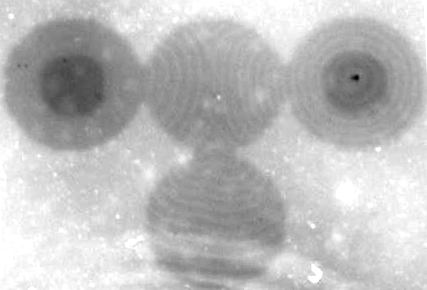}}
 \vspace{1cm}
\subfigure[]{\includegraphics[width=0.4\textwidth]{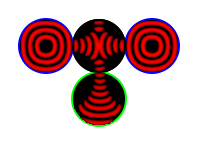}}
\caption{Example of laboratory work in progress.  (a) AND gate
  response to $(a,b)(1,1)\ra 1$ in real chemistry experiment. Fine
  wave fragments can be observed travelling from the input zones (dark
  areas, top left and right) and moving into the output disk as a
  result of collisions in the central disc. Disc diameter $\sim14 mm$,
  time duration from initiation to output cell bottom wall was
  $\sim510 s$. (b) AND gate simulation $(a,b)(1,1)\ra 1$ repeated for
  comparison from Sect.~\ref{section:logic_gates}.  Input discs are
  top left and right (blue rings), the output disc is bottom central
  (green ring).}
\label{figure:chemistry_and_gate}
\end{figure}

An example of some initial results from on-going laboratory work is
shown in Fig.~\ref{figure:chemistry_and_gate}a.  An identical
background image generated from the simulation software, where light
intensity is proportional to $\phi$ and is projected onto a thin layer
of silica gel containing a photo sensitive (Ru(bpy)$_3^{2+}$) catalyst
for the BZ reaction.  The gel is submerged in catalyst-free BZ
reagents (NaBrO$_3$, CH$_2$(COOH)$_2$, H$_2$SO$_4$ \&
BrMA).\footnote{Further details of chemistry and experimental
  apparatus are described in~\citep{Toth2009}.}  Waves are
continuously initiated by inserting silver colloidal particles onto
the gel surface at the center of the two input discs (dark centers,
top left and right).  Fine waves can be seen travelling out from each
input disc, colliding in the central disc and then expanding into the
output disk (bottom).  For comparison the result of the simulated AND
gate (Sect.~\ref{section:logic_gates}) is shown in the adjacent frame
(Fig.~\ref{figure:chemistry_and_gate}b).  The real chemistry image
(Fig.~\ref{figure:chemistry_and_gate}a) is a single image of multiple
wave initiations, whereas the simulation image
(Fig.~\ref{figure:chemistry_and_gate}b) is a composite of time lapse
images of a single wave initiation.

\section{Acknowledgements}
\label{section:Acknowledgements}

The work is part of the European project 248992 funded under 7th FWP
(Seventh Framework Programme) FET Proactive 3: Bio-Chemistry-Based
Information Technology CHEM-IT (ICT-2009.8.3). The authors wish to
acknowledge the support of the EPSRC grant number EP/E016839/1 for
support of Ishrat Jahan.  We would like to thank the project
coordinator Peter Dittrich and project partners Jerzy Gorecki and
Klaus-Peter Zauner for their inspirations and useful
discussions~\citep{neuneu2010}.

\bibliographystyle{./elsarticle-harv}

\renewcommand\refname{References}
\bibliography{./bz_bib}

\end{document}